\documentclass[prb,twocolumn,showpacs,preprintnumbers,amsmath,amssymb]{revtex4}
\usepackage{amsfonts}
\usepackage{graphicx}
\usepackage{dcolumn}
\usepackage{float}
\usepackage{bm}

\newcommand{\com}[1]{}

\begin{document}

\title{On stability of odd-frequency superconducting state}

\author{Dmitry Solenov, Ivar Martin, Dmitry Mozyrsky}
\affiliation{Theoretical Division (T-4), Los Alamos National
Laboratory, Los Alamos, NM 87545, USA}

\date{\today}

\begin{abstract}
Odd-frequency pairing mechanism of superconductivity has been investigated for several
decades. Nevertheless, its properties, including
the thermodynamic stability, have remained unclear. In
particular, it has been argued that the
odd-frequency state is thermodynamically unstable, has an
unphysical (anti-) Meissner effect, and thus can not exist as a homogeneous equilibrium phase.
We argue that this conclusion
is incorrect because it implicitly relies on the inappropriate assumption that
the odd-frequency superconductor can be described by an effective
Hamiltonian that breaks the particle conservation symmetry. We demonstrate that the
odd-frequency state can be properly described within the
functional integral approach  using non-local-in-time effective action.  Within the saddle point
approximation, we find that this phase is thermodynamically stable, exhibits ordinary Meissner effect, and
therefore can be realized as an equilibrium homogenous state of matter.  
\end{abstract}


\maketitle

Odd-frequency pairing mechanism was first introduced by
Berezinskii \cite{Berezinskii} in an attempt to describe the
A-phase of superfluid $^3$He.  Although, it was latter found that
in the case of superfluid $^3$He the pairing is odd is space
($p$-wave) rather than in time, the question weather the same-spin
fermions can develop anomalous correlations that are odd in
frequency  (``$p$-wave'' along the time
direction) remained open. The interest to this pairing mechanism
was revived almost two decades later in the context of
high-temperature superconductors by Balatsky and Abrahams,
\cite{BalatskyAbrahams1} who generalized it to the case of
opposite spin pairing. At the same time it has been argued that
the odd-frequency superconducting order parameter does not
correspond to a thermodynamically stable phase,\cite{heid1,cox}
and thus may not be realized as a homogenous equilibrium state of
matter.\cite{heid2,hf} On the other hand, the authors of
Ref.~\onlinecite{BelitzKirkpatrick} have shown that path integral
formulation of the problem can lead to the opposite conclusion. No
resolution of this contradiction have been proposed so far. In
this report we demonstrate that odd-frequency superconductor is
thermodynamically stable and exhibits ordinary Meissner effect. 

The argument on instability of the odd-frequency superconductor
can be summarized as follows~\cite{bal,heid1}. Thermodynamic
stability is determined by the change of free energy in the
ordered phase, relative to the disordered one.  Near the
second-order transition point  it  can be written
as\cite{agd,heid1}
\begin{equation}\label{GreenF}
\delta \Omega \sim -{1\over \beta}\sum_{\omega, {\bf q}}
{\Delta(\omega, {\bf q})\Delta^+(\omega, {\bf q})\over
\omega^2+\xi^2_{\bf q}}
\end{equation}
with gap-functions $\Delta(\omega, {\bf q})$ and $\Delta^+(\omega,
{\bf q})$ related to the anomalous (Matsubara) Green's functions
$F(\omega, {\bf q}) =\int d\tau e^{i\omega\tau}\langle {\rm
T}_\tau c_{\bf q}(\tau) c_{\bf -q}(0)\rangle$ and $F^+(\omega,
{\bf q}) =\int d\tau e^{i\omega\tau}\langle {\rm T}_\tau
c^\dag_{-\bf q}(\tau) c^\dag_{\bf q}(0)\rangle$ through a
self-consistency relation
\begin{equation}\label{self}
\Delta (\omega, {\bf q}) = \sum_{\omega^\prime, {\bf q}^\prime}
D(\omega - \omega^\prime, {\bf q} - {\bf
q}^\prime)F(\omega^\prime, {\bf q}^\prime),
\end{equation}
and an identical relation for $F^+$ and $\Delta^+$ . Here
$D(\omega, {\bf q})$ is irreducible interaction between
quasiparticles with energies $\xi_{\bf q}$ (we assume that $D$ is
real and even in both $\omega$ and ${\bf q}$), and $\beta$ is the
inverse temperature.  Here and in the following we consider
spinless fermions---extension to the usual two-spin species case
is straightforward and irrelevant for our purpose.

The sign of $\delta \Omega$ can be determined by analyzing the
relation between $\Delta(\omega, {\bf q})$ and $\Delta^+(\omega,
{\bf q})$. The non-zero averages $\langle {\rm T}_\tau c_{\bf
q}(\tau) c_{\bf -q}(0)\rangle$ and $\langle {\rm T}_\tau
c^\dag_{\bf q}(\tau) c^\dag_{\bf -q}(0)\rangle$ can only be obtained if  they are taken with respect to a state with
broken U(1)-symmetry (particle number conservation).  If an appropriate broken-symmetry mean-field Hamiltonian $H_{MF}$ exists, then
\begin{eqnarray}
\label{definition} F(\tau, {\bf q}) = (1/Z){\rm Tr} \left[ e^{ -
\beta H_{MF}} {\rm T}_\tau e^{\tau H_{MF}} c_{\bf q} e^{ - \tau
H_{MF}} c_{-{\bf q}}\right],
\\ \nonumber
F^+(\tau, {\bf q}) =
(1/Z){\rm Tr} \left[ e^{ - \beta H_{MF}} {\rm T}_\tau e^{\tau
H_{MF}} c^\dag_{-{\bf q}} e^{ - \tau H_{MF}} c^\dag_{{\bf
q}}\right],
\end{eqnarray}
where $Z = {\rm Tr}[e^{-\beta H_{MF}}]$. A straightforward
comparison shows that the two Green's functions are related as
\begin{equation}\label{relation}
F^\ast(\tau, {\bf q}) = F^+(\tau, {\bf q}), \ {\rm or} \
F^\ast(\omega, {\bf q}) = F^+(-\omega, {\bf q}).
\end{equation}
As a consequence of Eq.~(\ref{self}) the functions $\Delta(\omega,
{\bf q})$ and $\Delta^+(\omega, {\bf q})$ obey identical relation
and therefore the product $\Delta(\omega, {\bf q})\Delta^+(\omega,
{\bf q})$ in Eq.~(\ref{GreenF}) is negative definite for the
odd-frequency $\Delta(\omega, {\bf q})$ producing $\delta\Omega
>0$.  As the result, one is forced to conclude that the odd-frequency superconducting phase is
thermodynamically unstable.\cite{heid1} This conclusion can also
be reached if one uses the Green's functions obtained in original
work by Berezinskii.\cite{Berezinskii}  As a related issue, one
also finds an unphysical Meissner effect (i.e., with the negative
Meissner kernel).\cite{bal}  Later it was suggested that the
odd-frequency state might exist as an inhomogeneous phase, where
the order parameter is modulated at the microscopic
level,\cite{cox, hf, heid2} or might be a manifestation of some
composite (even-frequency) order
parameter.\cite{BalatskyAbrahams2}

In what follows, we show that relation~(\ref{relation}) holds only
for the {\it even}-frequency anomalous Green's function.
\textit{Odd}-frequency superconducting state changes this
relation, and Eq.~(\ref{relation}) is modified to $F^\ast(\tau,
{\bf q}) = - F^+(\tau, {\bf q})$. The problem with the reasoning outlined above is that it assumed existence of a mean-field Hamiltonian, $H_{MF}$.  However, to account for the retardation effects, which are essential in the
case of the odd-frequency state, effective Hamiltonian language,
e.g. Eq.~(\ref{definition}), is inappropriate.  In other words, $H_{MF}$ for odd-frequency superconductivity does not exist.  Instead, one should consider an effective action that is essentially non-local in time.

To study the superconducting phase, we represent the
partition function of the system is a functional
integral\cite{negele}
\begin{equation}
\label{partition} Z = \int {{{\cal D}}\bar \psi {{\cal D}}\psi
}{\cal D}\Delta^\ast {\cal D}\Delta e^{ - {\cal S}(\bar \psi,
\psi, \Delta^\ast, \Delta)},
\end{equation}
with
\begin{eqnarray}\label{action}
S\!&=&\!\!\int\!\!{dx_1} \bar\psi (x_1)\!\left[
{\partial_\tau\!\!+\! \hat\xi } \right]\! \psi
(x_1)\!+\!\!\int\!\!{dx_1dx_2} {|\Delta (x_1,x_2)|^2\over D(x_1 -
x_2)} 
\\ \nonumber
&+&\!\!\!\int\!\!{dx_1dx_2}\!\!\left[\Delta^\ast(x_1\!, x_2\!)\psi
(x_2\!)\psi (x_1\!)\!+\!\!\Delta (x_1\!,x_2\!) \bar\psi
(x_1\!)\bar \psi(x_2\!)\right]\!\!,
\end{eqnarray}
where $\bar\psi(x)$ and $\psi(x)$ are conjugate Grassmann
variables corresponding to the fermionic fields $\psi^\dag ({\bf
r})=\sum_{\bf q}e^{-i{\bf q r}}c^\dag_{\bf q}$ and $\psi({\bf
r})=\sum_{\bf q}e^{i{\bf q r}}c_{\bf q}$, with $x$ labeling both
spacial ${\bf r}$ and (Matsubara) time $\tau$  coordinates, and
$\hat\xi$  is the kinetic energy operator, $\hat\xi =
-\nabla^2_{\bf r}/(2m)-\mu$.  In Eqs.~(\ref{partition},
\ref{action}) we have introduced the pairing field $\Delta(x_1,
x_2)$ via the standard Hubbard-Stratonovich
transformation\cite{negele} by decoupling the interaction term
$\int {dx_1dx_2} D(x_1 - x_2)\bar \psi(x_1)\bar \psi (x_2)\psi
(x_2)\psi (x_1)$.  Note that up to this point, no approximation
has been made.

As pointed out earlier, the anomalous Green's functions can be
defined only with respect to a state with the broken $U(1)$ symmetry.
Indeed it is easy to see that the quantity
\begin{equation}\label{average}
\int {{{\cal D}}\bar \psi {{\cal D}}\psi }{\cal D}\Delta^\ast {\cal D}\Delta \ \psi (x)\psi
(x^\prime)  e^{ - {\cal S}(\bar \psi, \psi, \Delta^\ast, \Delta)}
\end{equation}
is identically zero:  after integration over the $\Delta$ fields
we average $\psi (x)\psi (x')$ with respect to the action
containing only products $\bar\psi \psi$ of Grassmann variables.
This average is nothing but $F(x-x')$ defined earlier in terms of
the time-ordered average, which is indeed zero in normal phase.
While in the normal phase, the partition function $Z$ of
Eq.~(\ref{partition}) is dominated by the vicinity of $\Delta=0$,
below certain temperature the situation changes: The primary
contribution to $Z$ comes from $|\Delta(x_1,x_2)|\neq 0$, which
signals spontaneous $U(1)$ symmetry breaking.   In addition, for
spinless electrons, the superconducting state breaks either
spatial parity (e.g. $p$-wave superconductor) or time reversal
symmetry (odd-frequency superconductor).  In the ordered state,
one can expand the action in the vicinity of the non-zero
saddle-point value of the order parameter $\Delta_{MF}(x_1-x_2)$,
with mean field approximation corresponding to further neglecting
the fluctuations around the saddle point. At the mean-field level,
the anomalous correlation functions ($F$ and $F^+$) can be
expressed as
\begin{eqnarray}\label{eq:F-pathIntegral-def}
F(\tau\!-\!\tau';{\bf{r}}\!-\!{\bf{r}}')\!=\!Z^{ -
1}_{MF}\!\!\int\!\! {{{\cal D}}\bar \psi {{\cal D}}\psi } \psi
({\bf{r}},\tau )\psi ({\bf{r}}',\tau ') e^{ - {\cal S}_{MF}},
\\ \label{eq:Fbar-pathIntegral-def}
F^+(\tau\!-\!\tau';{\bf{r}}\!-\!{\bf{r}}')\!=\!Z^{ - 1}_{MF}
\!\!\int\!\! {{{\cal D}}\bar \psi {{\cal D}}\psi } \bar \psi
({\bf{r}},\tau )\bar \psi ({\bf{r}}',\tau ')e^{ - {\cal S}_{MF}},
\end{eqnarray}
where
\begin{eqnarray}\label{action2}
{\cal S}_{MF} = &&\int {dx_1} \bar \psi (x_1)\left[ {\partial_\tau
+\hat\xi } \right]\psi (x_1)
\\ \nonumber
&&+\!\!\int\!\!{dx_1dx_2}\Delta^\ast_{MF}(x_1\!-\!x_2)\psi
(x_2\!)\psi(x_1\!)
\\ \nonumber
&&+\!\!\int\!\!{dx_1dx_2}\Delta_{MF}(x_1\!-x_2) \bar\psi
(x_1\!)\bar \psi(x_2\!),
\end{eqnarray}
and $\Delta_{MF}(x_1-x_2)$
is again defined by the self-consistency condition,
Eq.~(\ref{self}).

Now we are ready to determine the relation between $F$ and $F^+$.  Due to the long history of the problem, and since we believe that this relation is the root of the divergent claims about the fate of the odd-frequency superconductivity, we present here all technical details.
First, let us take the complex conjugate of $F$:
\begin{equation}\label{eq:cc-of-F}
F^*(\tau-\tau ';{\bf{r}}-{\bf{r}}')\!=\!Z^{ - 1}_{MF}\!\!\int\!\!
{{{\cal D}}\bar\psi {{\cal D}} \psi } \,\bar\psi
({\bf{r}}'\!,\tau ')\bar\psi ({\bf{r}},\tau )e^{ - {\cal
S}_{MF}^\ast },
\end{equation}
where
\begin{eqnarray}\label{eq:cc-of-L}
&&{\cal S}_{MF}^\ast  = \int {dx} \left\{ {\partial _\tau  \bar\psi (x)}
  \psi(x) + {\hat\xi\bar\psi (x)}
 \psi(x)  \right\}\ \ \ \ \ \
\\ \nonumber
&&+ \int {dx_1dx_2} \Big[\Delta_{MF} (x_1-x_2)\bar\psi (x_1)\bar\psi (x_2)
\\ \nonumber
&&+ \Delta_{MF}^\ast (x_1-x_2) \psi (x_2)\psi
(x_1)\Big].
\end{eqnarray}
Integrating the first two terms in Eq.~(\ref{eq:cc-of-L}) by parts
we obtain $\int {dx} \bar\psi (x)\left[ { - \partial _\tau   +
\hat\xi} \right] \psi(x)$. Then defining the new variables
according to $\bar\eta ({\bf{r}},\tau ) = \bar \psi ({\bf{r}},
- \tau )$ and $\eta({\bf{r}},\tau ) = \psi ({\bf{r}}, -
\tau )$, and changing $\tau\to-\tau$ in every integral in
Eq.~(\ref{eq:cc-of-L}), we obtain
\begin{equation}\label{eq:cc-of-F-1}
F^* (\tau  - \tau ';{\bf{r}}-{\bf{r}}')\!=\!Z^{ -
1}_{MF}\!\!\int\!\! {{\cal D}\bar\eta {\cal D}\eta} \bar\eta
({\bf{r}}'\!,-\tau ') \bar\eta ({\bf{r}},-\tau )e^{ - {\cal
\tilde S}_{MF} }
\end{equation}
with
\begin{eqnarray}\label{eq:Lbar-1}
&&{\cal \tilde S}_{MF} = \int {dx} \bar \eta (x)\left[ {\partial _\tau   + \hat\xi } \right]\eta (x)
\\ \nonumber
&&+ \int {dx_1dx_2}\Big[\Delta_{MF}^\ast (\tau_2 - \tau_1
,{\bf{r}}_1-{\bf{r}}_2)\eta (x_2)\eta (x_1)
\\ \nonumber
&&+ \Delta_{MF}(\tau_2 - \tau_1
,{\bf{r}}_1-{\bf{r}}_2)\bar \eta (x_1)\bar \eta (x_2)\Big]
\end{eqnarray}

For even-in-$\tau$ $\Delta(\tau, {\bf r})$, (e.g., \textit{p}-wave
for a single spin species case) we have ${\cal \tilde S}_{MF} =
{\cal S}_{MF}$ and therefore comparing Eqs.~(\ref{eq:cc-of-F-1},
\ref{eq:Lbar-1}) with Eqs.~(\ref{eq:Fbar-pathIntegral-def},
\ref{action2}) we recover Eq.~(\ref{relation}).

For odd-in-$\tau$ $\Delta(\tau, {\bf r})$, (e.g., \textit{s}-wave
with a single spin species) we see that by changing $\tau\to
-\tau$ a minus sign is generated in the last two terms in the RHS of
Eq.~(\ref{eq:Lbar-1}) as compared to Eq.~(\ref{eq:cc-of-L}).  This
difference can be removed by another change of variables $\eta ({\bf{r}},\tau )
\to i\eta ({\bf{r}},\tau )$, $\bar\eta ({\bf{r}},\tau ) \to  -
i\bar\eta ({\bf{r}},\tau )$, which is a simple gauge
transformation.   We
obtain again ${\cal \tilde S}_{MF} = {\cal S}_{MF}$.   However, an additional
factor (-1) now appears before the entire path integrals owning to the
fact that the quantity $\bar\eta\bar\eta$ transforms into ($-\bar\eta\bar\eta$) as a result of the last gauge transformation.  Therefore comparing Eqs.~(\ref{eq:cc-of-F-1},
\ref{eq:Lbar-1}) with Eqs.~(\ref{eq:Fbar-pathIntegral-def},
\ref{action2}) we finally obtain that in the odd-frequency case we have
the relation
\begin{equation}\label{relation2}
F^\ast(\tau-\tau', {\bf r}-{\bf r}') = -F^+(\tau-\tau', {\bf
r}'-{\bf r})
\end{equation}
or
\begin{equation}\label{relation3}
F^\ast(\omega, {\bf q}) = F^+ (\omega, {\bf q}),
\end{equation}
Obviously Eq.~(\ref{relation3}) holds for the even-frequency case
as well --- in which case $F^+(-\omega, {\bf q})$ in
Eq.~(\ref{relation}) can be replaced by $F^+(\omega, {\bf q})$ for
even-frequency $F$. As a result, contrary to the conclusion of
Refs.~\onlinecite{heid1,cox,hf,heid2}, the product $\Delta(\omega,
{\bf q})\Delta^+(\omega, {\bf q})$ in Eq.~(\ref{GreenF}) is
positive definite both for odd- and even-frequency pairing and
therefore $\delta\Omega <0$. The same conclusion can be reached by
directly analyzing the mean-field free energy, given by
Eqs.~(\ref{partition}, \ref{action}), as the system undergoes the
phase transition. Thus we conclude that the odd-frequency
superconducting phase has free energy lower than that of the
normal phase.  The magnitude of the order parameter has to be
determined from the self-consistency condition and is non-zero
below the superconducting transition temperature.

It can be also verified that due to relation (\ref{relation3})
odd-frequency superconducting phase has a positive Meissner kernel
and therefore a physically meaningful Meissner effect. The
supercurrent and the vector potential are related as
$\mathbf{j}(\mathbf{k})=-\mathcal{K}(\mathbf{k})\mathbf{A}(\mathbf{k})$,
where the (Meissner) kernel $\mathcal{K}$ is expressed\cite{agd}
as
\begin{eqnarray}\nonumber
\mathcal{K}(\mathbf{k})=\frac{Ne^2}{m} &+&
\frac{2e^2}{m^2\beta}\sum_{\omega}\int\!\!\!\frac{d\mathbf{p}}{(2\pi)^3}
\mathbf{p}^2
\left[\mathcal{G}(\omega,\mathbf{p}_+)\mathcal{G}(\omega,\mathbf{p}_-)
\right.
\\\label{eq:Q-kernel}
&+&\left. F(\omega,\mathbf{p}_+)F^+(\omega,\mathbf{p}_-)\right].
\end{eqnarray}
Here $\mathbf{p}_\pm = \mathbf{p}\pm \mathbf{k}/2$. The Green's
functions can be easily obtained from
Eqs.~(\ref{eq:F-pathIntegral-def}, \ref{eq:Fbar-pathIntegral-def},
\ref{action2}). We have
\begin{eqnarray}\label{eq:F}
F(\omega, {\bf q}) = \frac{2\Delta_{MF}(\omega, {\bf
q})}{\omega^2+\xi(\mathbf{q})^2+4|\Delta_{MF}(\omega,\mathbf{q})|^2}
\end{eqnarray}
\begin{eqnarray}
\label{eq:F-1} F^+(\omega, {\bf q}) =
\frac{2\Delta^\ast_{MF}(\omega, {\bf
q})}{\omega^2+\xi(\mathbf{q})^2+4|\Delta_{MF}(\omega,\mathbf{q})|^2}
\end{eqnarray}
\begin{eqnarray}\label{eq:G}
\mathcal{G}(\omega, {\bf q}) =
\frac{i\omega+\xi(\mathbf{q})}{\omega^2+\xi(\mathbf{q})^2+4|\Delta_{MF}(\omega,\mathbf{q})|^2}
\end{eqnarray}
Note that the form of Eqs.~(\ref{eq:F}, \ref{eq:F-1}) is
consistent with Eq.~(\ref{relation3}), not with
Eq.~(\ref{relation}). Had we used Eq.~(\ref{relation}), we would
have obtained $\Delta^\ast_{MF} (-\omega, {\bf q})$ in
Eq.~(\ref{eq:F-1}) as well as $-|\Delta(\omega,\mathbf{q})|^2$ in
the denominators of Eqs.~(\ref{eq:F}, \ref{eq:F-1}, \ref{eq:G}).

As usual Eq.~(\ref{eq:Q-kernel}) is divergent and we regularize it
by subtracting $\mathcal{K}(\mathbf{k})$ for $\Delta_{MF}(\omega,
{\bf q})=0$ (obviously $\mathcal{K}(\mathbf{k})=0$ in normal
phase).\cite{agd} In the long wave-length limit and for
$\Delta_{MF}(\omega, {\bf q})$ independent of ${\bf q}$ (i.e., for
pairing in $s-wave$ channel) we obtain
\begin{equation}\label{eq:Q-result}
\mathcal{K}(\mathbf{k}\to 0)=\frac{\pi N e^2}{m\beta}
\sum_{\omega} \frac{4|\Delta_{MF}(\omega)|^2
}{(\omega^2+4|\Delta_{MF}(\omega)|^2)^{3/2}}.
\end{equation}
This equation obviously is positive definite. Note that if we had
used Eq.~(\ref{relation}) (which is invalid as we argue above), we
would have obtained $\Delta^+(\omega, {\bf q})\Delta(\omega, {\bf
q}) =  \Delta^\ast(-\omega, {\bf q})\Delta(\omega, {\bf q})$ in
the numerator in the RHS of Eq.~(\ref{eq:Q-result}) and thus
negative Meissner kernel for the odd-frequency case.

While possessing similar electromagnetic properties (i.e., the Meissner effect)
to its familiar even-frequency counterpart, the odd-frequency superconductor is
expected to differ from the even-frequency one in several important aspects.
Here we mention just some of them qualitatively.  The equal spin pairing
considered above, leads to a gapless superconductor, with an isotropic
($s$-wave) electronic spectral function.  The only other known example where
this can happen is the $s$-wave superconductor with a relatively high
concentration of magnetic impurities; however, for odd-frequency
superconductor this would occurs even in the clean case.  The
odd-frequency anisotropic $s$-wave superconductor can readily exceed the
Pauli paramagnetic limit, and can show very little change in magnetic
susceptibility (Knight shift) across the superconducting transition.
These and other properties, in particular the manifestations of the
time-reversal symmetry braking in such superconductors, are attractive
direction for future detailed theoretical investigations.

We acknowledge valuable and stimulating discussions with E. Abrahams and A. Balatsky. This
work  is supported by the US DOE.


\end{document}